\documentclass[a4paper,usenatbib]{mnras}

\usepackage{footnote}
\usepackage{amsmath}
\usepackage{amssymb}
\usepackage{xcolor}
\usepackage{graphicx}
\usepackage{subfigure}
\usepackage{pdflscape}
\usepackage{url}

\DeclareGraphicsExtensions{.pdf,.png,.jpg}

\title{Weak CS Emission in an Extremely Metal-poor Galaxy DDO~70}

\author[K. Du et al.]{
  Kaiyi Du$^{1,2}$,
  Yong Shi$^{1,2}$ \thanks{Email: yong@nju.edu.cn},
  Zhi-Yu Zhang$^{1,2}$ \thanks{Email: zzhang@nju.edu.cn},
  Junzhi Wang$^{3}$,
  Yu Gao$^{4,5}$
  \\
  $^{1}$School of Astronomy and Space Science, Nanjing University, Nanjing 210093, China \\
  $^{2}$Key Laboratory of Modern Astronomy and Astrophysics (Nanjing University), Ministry of Education, Nanjing 210093, China \\
  $^{3}$Shanghai Astronomical Observatory, Chinese Academy of Sciences, 80 Nandan Road, Shanghai 200030, China\\
  $^{4}$Department of Astronomy, Xiamen University, Xiamen, Fujian 361005, China\\
  $^{5}$Purple Mountain Observatory, Chinese Academy of Science, Nanjing 210008, China
}

\begin{document}
\label{firstpage}
\pagerange{\pageref{firstpage}--\pageref{lastpage}}
\maketitle

\begin{abstract}
In most galaxies like the Milky Way, stars form in clouds of molecular gas. Unlike the CO emission that traces the bulk of molecular gas, the rotational transitions of HCN and CS molecules mainly probe the dense phase of molecular gas, which has a tight and almost linear relation with the far-infrared luminosity and star formation rate. However, it is unclear if dense molecular gas exists at very low metallicity, and if exists, how it is related to star formation. In this work, we report ALMA observations of the CS $J$=5$\rightarrow$4 emission line of DDO~70, a nearby gas-rich dwarf galaxy with $\sim7\%$ solar metallicity. We did not detect CS emission from all regions with strong CO emission. After stacking all CS spectra from CO-bright clumps, we find no more than a marginal detection of CS $J$=5$\rightarrow$4 transition, at a signal-to-noise ratio of $\sim 3.3$. This 3-$\sigma$ upper limit deviates from the $L^\prime_{\rm CS}$-$L_{\rm IR}$ and $L^\prime_{\rm CS}$-SFR relationships found in local star forming galaxies and dense clumps in the Milky Way, implying weaker CS emission at given IR luminosity and SFR. We discuss the possible mechanisms that suppress CS emission at low metallicity.
\end{abstract}


\section{Introduction}\label{sec:intro}

Molecular clouds are the main sites of star formation in most galaxies. Studies of Galactic star formation indicate that stars, especially the massive stars, mainly form in the dense cores of giant molecular clouds \citep{McKee2007, Zinnecker2007}. The raw material reservoir of star formation is provided by molecular gas \citep{Bolatto2013}, except the uncertainty of the very first generations of stars.  

Low-$J$ CO lines are the best tracer of the bulk of molecular gas because of its small dipole moment, low upper energy levels, and high abundance. CO lines serve as a powerful tool to understand the role of gas in driving star formation across the Universe \citep{Kennicutt1989, Genzel2010, Shi2018}. However, CO lines are not suitable probes of regions with gas densities ten times higher than the average conditions in giant molecular clouds \citep{Gao2004}, where the star-forming gas locates.

Molecules with high dipole moments, e.g., HCN, CS, HCO$^{+}$, require much higher H$_{2}$ volume densities to excite, therefore their rotational transitions can trace H$_{2}$ gas at densities $\rm n_{H}\gtrsim3\times10^{4}cm^{-3}$ \citep{Gao2004, Bergin2007}. Studies have found tight and linear correlations between luminosities of HCN $J$=1$\rightarrow$0 and IR emission \citep{Gao2004, Gao2007, Krips2008, Chen2015, Privon2015}. \citet{Wu2010} and \citet{Wang2011} show linear correlations between the IR luminosity and that of CS $J$=5$\rightarrow$4 in massive Galactic dense cores and spiral galaxies, respectively. \citet{Zhang2014} found linear correlations between the IR luminosity and the dense gas tracers including CS $J$=7$\rightarrow$6 and HCN $J$=4$\rightarrow$3, and a slightly super-linear correlation for HCO$^{+}$ $J$=4$\rightarrow$3.

On the other hand, some studies \citep{Kauffmann2017, Shimajiri2017} found that those so-called dense gas tracers, e.g. HCN, HCO$^{+}$, do not particularly trace dense gas regions. Recent studies, \citet{Nguyen-Luong2020} and \citet{Evans2020}, also found the a large fraction of the total luminosity of HCN $J$=1$\rightarrow$0 and HCO$^{+}$ $J$=1$\rightarrow$0 emission arises in less dense molecular gas regions in inner Galactic molecular clouds and M17 clouds. At this point, the CS emission may also arises from more extended and less dense regions. Observations with high resolution can provide samples for studying the complex cases of excitation of CS molecules in interstellar medium.

Few studies have extended the study of dense gas tracers to galaxies at very low metallicities, where even CO emission is very faint \citep{Taylor1998, Leroy2005, Leroy2007, Cormier2014, Shi2015, Shi2016}. Extremely metal-poor galaxies, with $Z\, <\, 10\%Z_\odot$, are chemically unevolved, providing the best local analogues for primordial galaxies in the early Universe \citep{Shi2014, Hunt2014}. The quest for dense gas emission in these galaxies thus offer insights into star formation in the nascent galaxies. 

With the IRAM 30-m telescope, \citet{Shi2016} detected CO $J$=2$\rightarrow$1 in an extremely metal-poor galaxy, DDO~70, and provided a direct evidence for the existence of molecular gas in galaxies of such low metallicity. Our follow-up ALMA observation further resolved the CO emission into a few clumpy structures at a spatial resolution of 1.4 pc \citep{Shi2020}. The same observation also covers the CS $J$=5$\rightarrow$4 transition, offering an opportunity to explore the complex cases of CS excitation at such extremely low metallicity.

\section{Data} \label{sec:data}

DDO~70 is an extremely metal-poor galaxy with a gas-phase oxygen abundance of $\rm 12+log(O/H)=7.53$ \citep{Kniazev2005}, $\sim7\%$ of the Solar value (the Solar abundance is $\rm 12+log(O/H)=8.66$ \citep{Asplund2009}). This galaxy is at a distance of 1.38 Mpc \citep{Tully2013} and a redshift of 0.001004 \citep{Springob2005}.

As detailed in \citet{Shi2020}, our ALMA observations consist of two runs on 11 Nov. 2016 and 03 Aug. 2017, with the C40-4 and C40-7 array configurations at their best angular resolutions of 0.10$''$ and 0.35$''$, respectively. The on-source exposure times were 37 minutes and 63 minutes, respectively. After combining two data sets together, the final achieved r.m.s noise level is $\rm \sim 1.0\,mJy/beam$ at a velocity resolution of $\rm 0.32\, km/s$. Natural weighting was adopted in the {\sc clean} process to optimize sensitivity. The final synthesized beam is $0.22''\,\times\,0.19''$.

\begin{figure}
  \centering
  \includegraphics[width=0.9\linewidth]{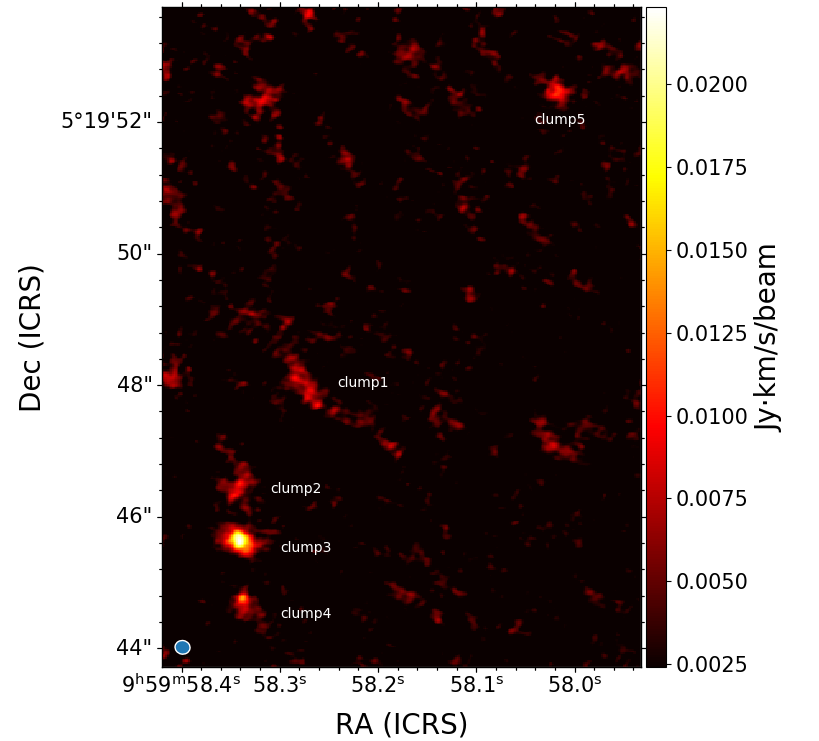}
\caption{Velocity-integrated flux density map of CO $J$=2$\rightarrow$1 of the star-forming region in DDO~70. We label the five clumps with significant CO emission ( integrated flux $\rm S/N>4$ ) in this map. The synthesized beam is $0.22'' \times\,0.19''$, shown in the lower-left corner.}
  \label{fig:1}
\end{figure}

We used CASA 5.4.0 \citep{CASA2007} to extract the CS $J$=5$\rightarrow$4 spectra from five clumps with significant CO emission identified in \citet{Shi2020}, which are shown in Figure \ref{fig:1}. We did not detect CS $J$=5$\rightarrow$4 in any of the CO clumps. Therefore, to improve the signal-to-noise ratio level, we further stack the CS spectra from all CO-bright clumps. 

To minimize noise contamination, we convolve the CO moment-0 map with twice the synthesized beam, make a mask of 3-$\sigma$ flux threshold, and then extract both CO and CS spectra from the data with original resolution. The total CO flux summed up from all five clumps accounts for $\sim 80\%$ of the emission seen by IRAM 30-m telescope \citep{Shi2020}, suggesting that the extended CO and CS emission is not systematically missed in our ALMA data.

\begin{figure}
  \centering
  \includegraphics[width=1.0\linewidth]{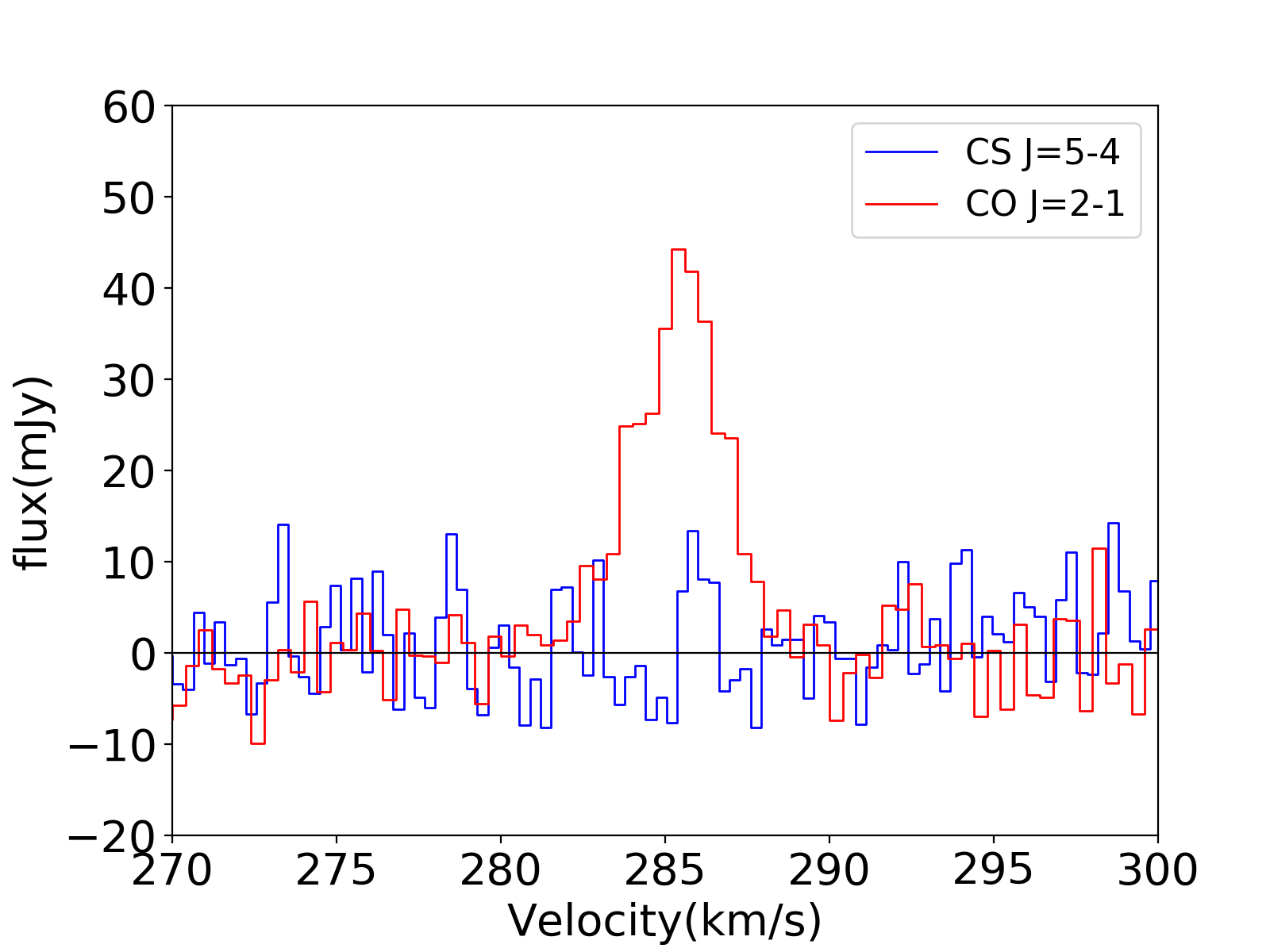}
\caption{Stacked spectra of CO $J$=2$\rightarrow$1 and CS $J$=5$\rightarrow$4 in DDO~70. The spectra are extracted from five individual CO-bright clumps detected with ALMA. The red line presents the stacked spectrum of CO $J$=2$\rightarrow$1, and the blue line shows the stacked spectrum of CS $J$=5$\rightarrow$4. The CS emission has an $\rm S/N \sim 3.3$ and a much narrower line width compared to the CO spectrum.}
  \label{fig:2}
\end{figure}

Figure \ref{fig:2} shows the stacked spectra of CO $J$=2$\rightarrow$1 and CS $J$=5$\rightarrow$4 transition. The CO $J$=2$\rightarrow$1 emission is detected at very high signal-to-noise ratio, with a luminosity of $\rm \sim156.2\,K\,km\,s^{-1}\,pc^{2}$, according to Equation \ref{eq.1} of \citet{Solomon2005}:

\begin{equation}
L^{'}=3.25\times 10^{7}(1+z)^{-3}f_{\rm obs}^{-2}S_{\rm CO}D_{\rm L}^{2},
     \label{eq.1}
\end{equation}

, which converts the velocity-integrated flux density, $S_{\rm CO}$ (measured in $\rm Jy\,km\,s^{-1}$), to luminosity in units of $\rm K\,km\,s^{-1}\,pc^{2}$. We adopt the Full Width Zero Intensity (FWZI) line width for CO $J$=2$\rightarrow$1. $D_{\rm L}$ is the luminosity distance at redshift $z$ and $f_{\rm obs}$ is the observing frequency in GHz. As shown in Figure \ref{fig:2}, the stacked CS $J$=5$\rightarrow$4 emission line shows a marginal detection feature with a signal to noise ratio (S/N) of $\sim 3.3$ in the integrated flux. This marginal feature is within the velocity range of the CO $J$=2$\rightarrow$1 line, and shows a much narrower line width.

Similar to what was found in \citet{Lapinov1998}, the line width of CS $J$=5$\rightarrow$4 emission line is much narrower than that of CO $J$=1$\rightarrow$0. Since the line widths of CO $J$=2$\rightarrow$1 and CO $J$=1$\rightarrow$0 are nearly the same, it seems reasonable that CS $J$=5$\rightarrow$4 emission of DDO~70 has such narrow line width. Because the detection is no more than marginal, we used the 3-$\sigma$ upper limit of CS $J$=5$\rightarrow$4 luminosity in the study, which is $\rm \sim10.9\,K\,km\,s^{-1}\,pc^{2}$.

\section{Results} \label{sec:result}
\begin{figure}
  \centering
  \subfigure{ \includegraphics[width=1.0\linewidth]{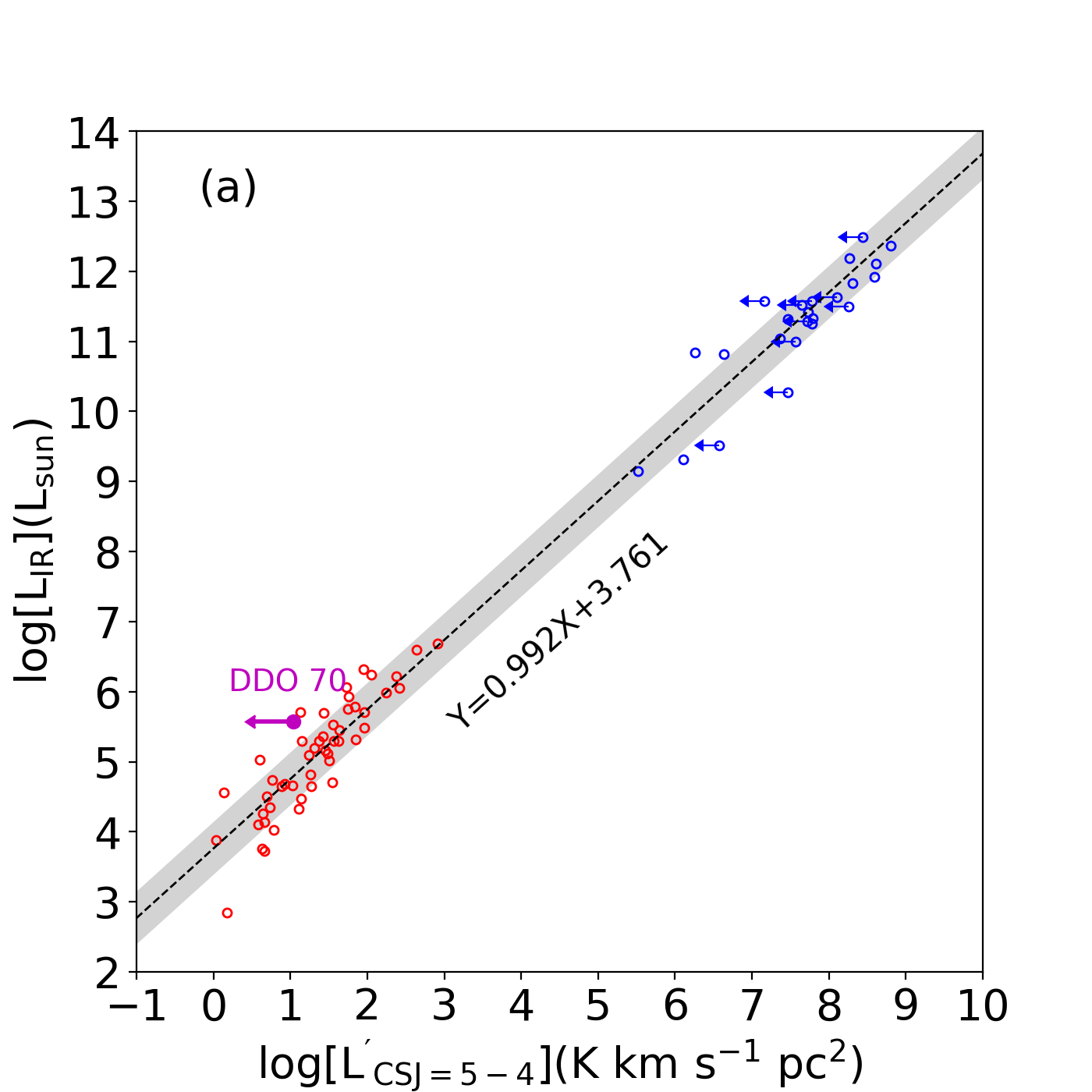}\label{fig:3a}}
  \subfigure{\includegraphics[width=1.0\linewidth]{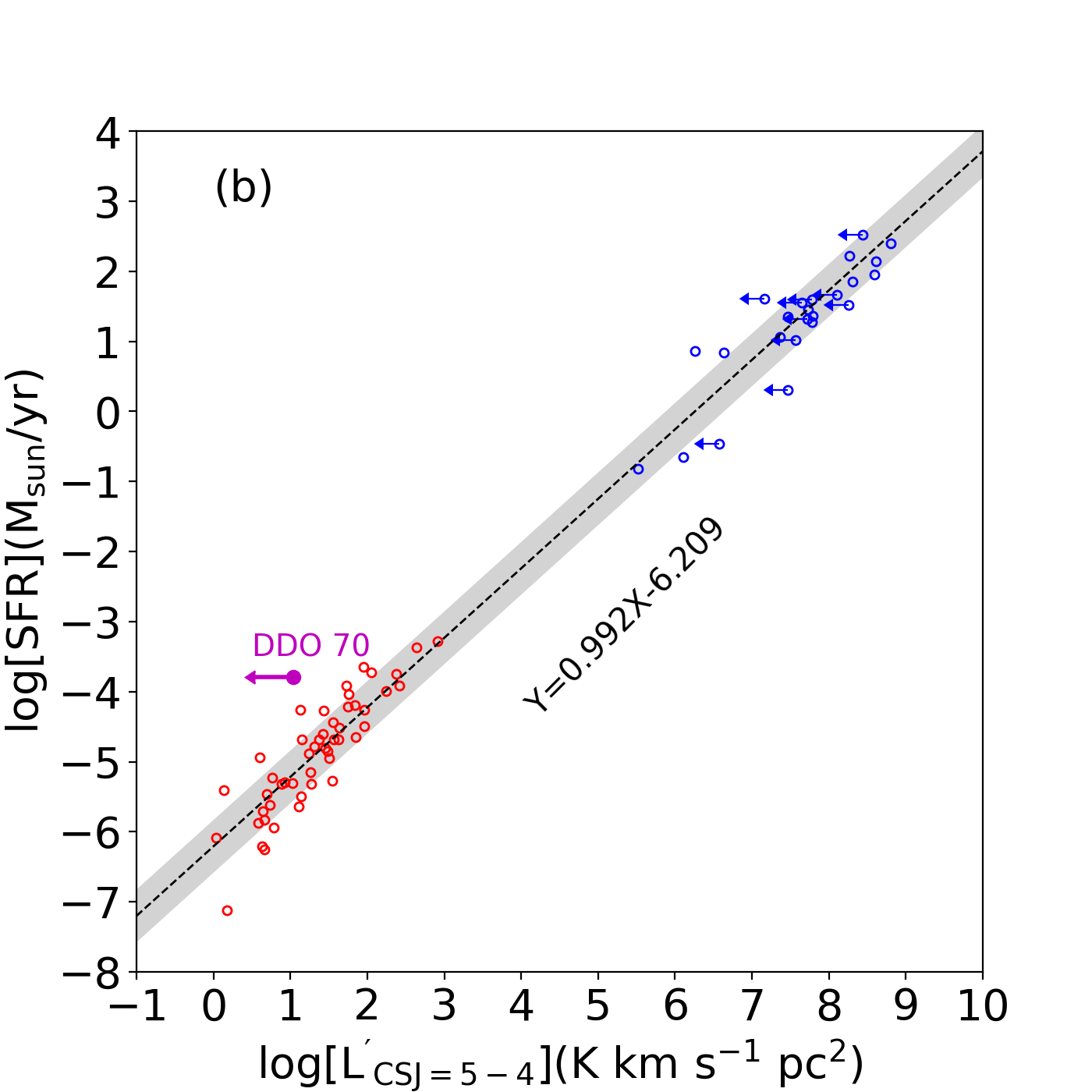}\label{fig:3b}}
  \caption{
 (a) $L_{\rm IR}$ as a function of $L^\prime_{\rm CS J=5-4}$ of galaxies in \citet{Wang2011}(open blue circles) and Galactic dense cores in \citet{Wu2010}(open red circles). We fit a correlation (black dashed line) between the luminosities of CS $J$=5$\rightarrow$4 and IR from the two sample, with upper limits excluded \citep{Wang2011}. The grey region shows the 1-$\sigma$ scatter bounds of this best-fit model. The upper limit luminosity of CS $J$=5$\rightarrow$4 of DDO~70 lies above the 1-$\sigma$ scatter upper bound. (b) SFR as a function of $L^\prime_{\rm CS J=5-4}$ in the same sample. The black dashed line shows the fitted correlation. The grey region shows the 1-$\sigma$ scatter bounds of the best fit.
  }
\end{figure}

As shown in Figure \ref{fig:3a}, we compiled both extra-galactic and Galactic measurements of CS $J$=5$\rightarrow$4 from \citet{Wang2011} and \citet{Wu2010}. The combined data show a correlation between the luminosities of CS $J$=5$\rightarrow$4 and IR emission. We refit all data points with a straight line, excluding upper limits. The derived best-fit line is consistent with the empirical logarithmic relationship found in previous studies. The 3-$\sigma$ upper limit of CS $J$=5$\rightarrow$4 of DDO~70 is located above the 1-$\sigma$ upper bound of the linear correlation, which may imply a deficit of CS molecules at such low metallicity. Assuming that the IR luminosities represent the star-formation rates (SFR) in galaxies, we converted the IR luminosities of the sample in \citet{Wang2011} and \citet{Wu2010} into the SFR using the following equation in \citet{Kennicutt1998}:

\begin{equation}
SFR(M_{\odot}/yr)=2.8\times 10^{-44}L(IR)(erg \cdot s^{-1}),
     \label{eq.2}
\end{equation}

where the initial mass function (IMF) is adapted from \citet{Chabrier2003}.

The amount of dust in DDO~70 is expected to be much less than that of normal galaxies, and the SFR calculated only with IR emission would underestimate the total SFR \citep{Hayward2014}. Therefore, we adopted the UV and IR combined SFR of $\rm 1.6\times10^{-4}\,M_{\odot}/yr$ within the ALMA aperture of DDO~70, as detailed in \citet{Shi2020}. Figure \ref{fig:3b} shows the correlation between $L^\prime_{\rm CS}$ and the updated SFR, where DDO~70 also shows a larger offset from this relationship.

\begin{figure}
  \centering
  \includegraphics[width=1.0\linewidth]{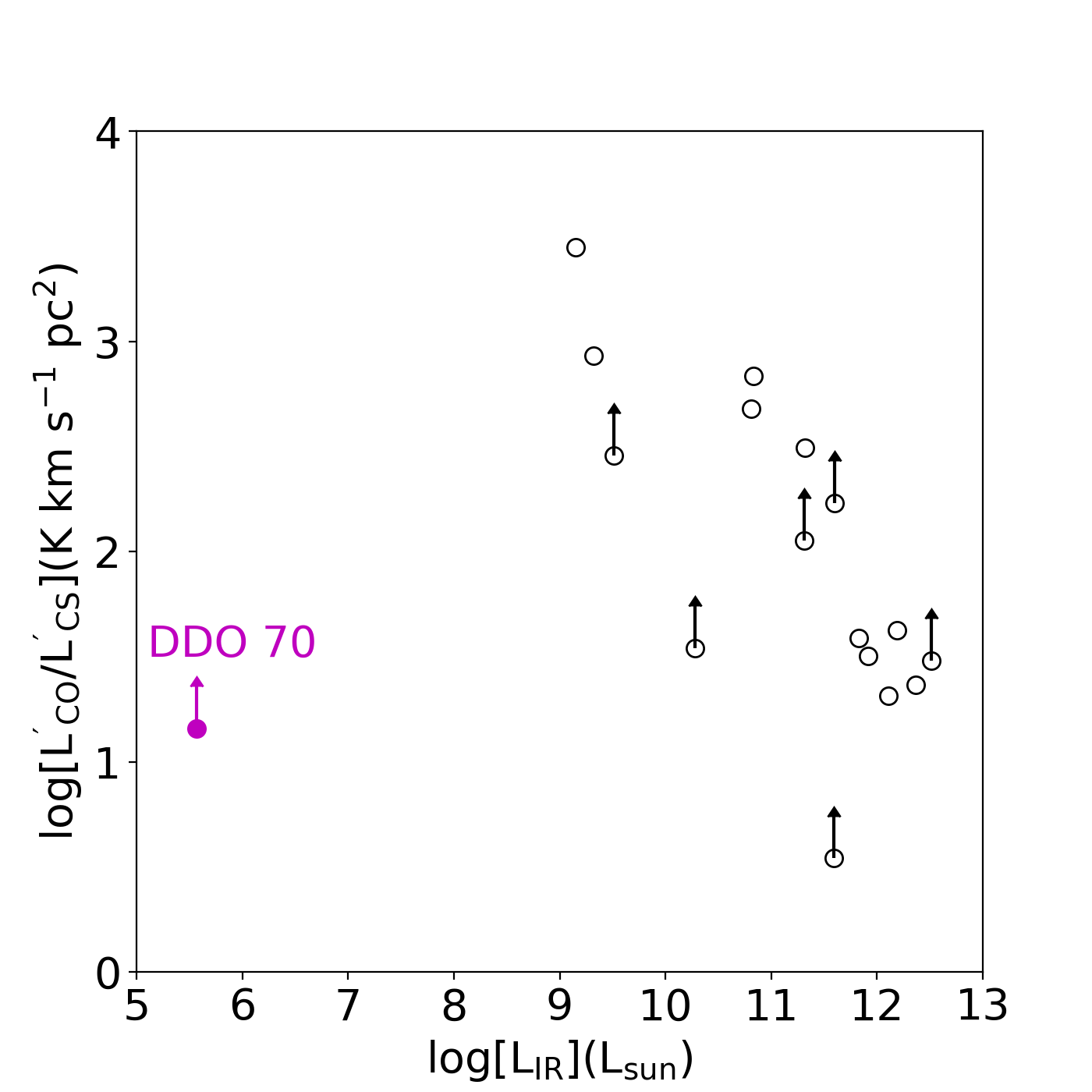}
\caption{
 Ratios of line luminosities of CO $J$=1$\rightarrow$0 to CS $J$=5$\rightarrow$4 as a function of IR luminosity. Black circles show galaxies from \citet{Wang2011}, for which the CO $J$=1$\rightarrow$0 data are taken from \citet{Gao2004a}. DDO~70 is plotted as the filled magenta circle.
 }
  \label{fig:4}
\end{figure}

We also calculated the ratio of the line integrated luminosity of CO $J$=1$\rightarrow$0 and CS $J$=5$\rightarrow$4 of DDO~70. Molecular clouds in Milky Way have an average CO $J$=2$\rightarrow$1/CO $J$=1$\rightarrow$0 line ratio from 0.89 for galactic nuclei \citep{Braine1992} to 0.6 in the Solar neighborhood \citep{Hasegawa1997}. The average CO $J$=2$\rightarrow$1/CO $J$=1$\rightarrow$0 ratios are $\sim0.7$ and $\sim0.89$ in nearby early-type galaxies \citep{Baldi2015} and spiral galaxies \citep{Braine1992}, respectively. Because dwarf galaxies have warm interstellar medium (ISM) indicated by their relatively higher dust temperature \citep{Shi2014, Zhou2016}, we assumed that CO $J$=2$\rightarrow$1/CO $J$=1$\rightarrow$0 line ratio is $\sim1$, under optically thick and local thermodynamic equilibrium conditions. Figure \ref{fig:4} shows the ratio between CO $J$=1$\rightarrow$0 and CS $J$=5$\rightarrow$4 against the IR luminosity of DDO 70 along with data from \citet{Wang2011} and \citet{Gao2004a}. The lower limit of CO $J$=1$\rightarrow$0/CS $J$=5$\rightarrow$4 of DDO~70 is consistent with IR bright galaxies.

\section{Discussion} \label{sec:dis}
We found no more than marginal detection of CS $J$=5$\rightarrow$4 in DDO~70. The 3-$\sigma$ upper limit of CS $J$=5$\rightarrow$4 in DDO 70 locates at above the best fit of the $L^\prime_{\rm CS}$-$L_{\rm IR}$ and $L^\prime_{\rm CS}$-SFR relations. This suggests that the CS emission in DDO~70 is very weak for the given IR luminosity compared to spiral galaxies and the Milky Way molecular clouds, and DDO~70 deviates from the $L^\prime_{\rm CS}$-$L_{\rm IR}$ relation. One possibility is that the molecular gas in DDO~70 may not have sufficient density to efficiently excite the CS lines, which is consistent with the finding of \citet{Shi2020}. Because of inefficient gas cooling and weak turbulence at very low metallicity, the large clumps cannot fragment into small ones and the average gas density of a clump is lower (see Figure 8 in \citet{Shi2020}), leading to inefficient star formation \citep{Shi2014}. As a result, a molecular cloud at low metallicity has on average lower densities, resulting in weaker CS emission.

Another explanation for weak CS emission in DDO~70 is that the abundances of gas-phase carbon and sulfur are too low due to the extremely low metallicity, leading to weak CS emission. \citet{Braine2017} found weak HCN emission in low-metallicity galaxies but with relatively normal CO and $\rm HCO^{+}$ emission. This weak HCN emission is likely due to the relatively low N/O abundance ratio in a low oxygen abundance environment. However, the relative abundance of $\alpha$ elements, including sulfur, is roughly constant for different oxygen abundance \citep{vanZee2005}. So the dominant cause for the weak HCN line in \citet{Braine2017} may not apply to our CS study. 

On the other hand, as the metallicity declines, the total amount of the dust decreases, which lowers the IR luminosity as well. Besides, as shown in \citet{Ledoux2003}, \citet{Wolfe2005} , \citet{Meiring2009}, \citet{DeCia2013}, and \citet{DeCia2016}, the strength of dust depletion effect decreases when metallicity declines, meaning a much smaller fraction of metals depleted onto dust grain surface at low metallicity than at high metallicity. Therefore, the IR luminosity may decrease more rapidly than that of CS $J$=5$\rightarrow$4 toward the low end of the range of metallicity. DDO~70 is expected to be below the relationship, which is inconsistent with the observation.

One more possible explanation for the lack of CS emission is the difficulty for carbon and sulfur to stay in the form of CS molecules. The interstellar radiation field could destroy CS molecules efficiently at low metallicities. When the metallicity declines, the amount of dust decreases rapidly, and the dust extinction $A_{\rm V}$ decreases. \citet{Lee2015, Lee2018} show that the dust shielding protects CO molecules from photodissociation efficiently within the extinction threshold, below which the CO molecules are largely destroyed. For typical molecular clouds in Milky Way, the dust extinction can reach a few or tens of magnitude \citep{Lee2018}. But for the extremely low-metallicity conditions like DDO~70, dust protection is much less. As studied in \citet{Shi2020}, the extinctions for CO-bright clumps are between 1 and 2.5 mag, estimated from the $I_{\rm CO}$-$A_{\rm V}$ relation for SMC \citep{Lee2018}. \citet{Heays2017} suggests that the photodissociation rate of molecules increases rapidly with decreasing extinction, when the extinction is below two magnitude. Thus, for DDO~70, few CS molecules can survive against the strong photodissociation effect, even within dense regions.
  
In the summary, the lack of carbon and sulfur elements, the decrease of dust, and the strong photodissociation field could all contribute to the deviation of DDO~70 from the $L^\prime_{\rm CS}$-$L_{\rm IR}$ and $L^\prime_{\rm CS}$-SFR correlations. In the end, it is still unknown that whether these two effects can fully contribute to such low luminosity and whether there exists other dominant physical mechanisms.

\section{Conclusion}\label{sec:con}
We present the ALMA observations of CS $J$=5$\rightarrow$4 in DDO~70, an extremely metal-poor galaxy with the existence of molecular gas CO. We detected strong CO line emission in five clumps but none for CS $J$=5$\rightarrow$4 at the same locations. After stacking all CS spectra from the five CO-bright clumps, we find no more than a marginal signal of CS $J$=5$\rightarrow$4 emission, with S/N $\sim 3.3$. If we adopt the upper limit of CS emission of DDO 70, it slightly deviates from the linear $L^\prime_{\rm CS}$-$L_{\rm IR}$ and $L^\prime_{\rm CS}$-SFR correlations by star-forming galaxies and Galactic dense cores. 

Weak CS emission may be result of on average lower density of a molecular cloud at very low metallicity. The the low CS $J$=5$\rightarrow$4 luminosity may be affected by the low gas-phase carbon and sulfur abundance too. Rapid photodissociation in reducing the abundance of the species of CS molecules may also play an important role.

\section{Acknowledgement}
We sincerely thank the referee for helping to significantly improve the paper. K.D. and Y.S. acknowledges the support from the National Key R\&D Program of China (No. 2017YFA0402704, No. 2018YFA0404502), the National Natural Science Foundation of China (NSFC grants 11825302, 11733002 and 11773013). Y. S. thanks the support from the Tencent Foundation through the XPLORER PRIZE.

YG's research is supported by National Key Basic Research and Development Program of China (grant No. 2017YFA0402704), National Natural Science Foundation of China (grant Nos. 11861131007, U1731237, 11420101002), and Chinese Academy of Sciences Key Research Program of Frontier Sciences (grant No. QYZDJSSW-SLH008).

This work also benefited from the International Space Science Institute (ISSI/ISSI-BJ) in Bern and Beijing, thanks to the funding of the team “Chemical abundances in the ISM: the litmus test of stellar IMF variations in galaxies across cosmic time” (Principal Investigator D.R. and Z-Y.Z.).

This paper makes use of the following ALMA data: ADS/JAO.ALMA\#2016.1.00359.S. ALMA is a partnership of ESO (representing its member states), NSF (USA) and NINS (Japan), together with NRC (Canada) and NSC and ASIAA (Taiwan) and KASI (Republic of Korea), in cooperation with the Republic of Chile. The Joint ALMA Observatory is operated by ESO, AUI/NRAO and NAOJ.

\bibliography{DDO70.bib}
\bibliographystyle{mnras}

\end{document}